\documentclass[useAMS,usenatbib]{mn2e}
\usepackage{amsmath}
\usepackage{graphicx}
\usepackage{txfonts}
\usepackage{subfigure}
\usepackage{color}
\usepackage{amstext}

\title[Observability of a nitrogen dominated atmosphere for 55 Cnc e]{Observability of molecular species in a nitrogen dominated atmosphere for 55 Cancri e}
\author[Y. Miguel]{Y. Miguel$^{1}$\thanks{E-mail:
ymiguel@strw.leidenuniv.nl}\\
$^{1}$Leiden Observatory, University of Leiden, Niels Bohrweg 2, 2333CA Leiden, The Netherlands\\
}
\begin{document}

\pagerange{\pageref{firstpage}--\pageref{lastpage}} \pubyear{2002}

\maketitle

\label{firstpage}

\begin{abstract}
One of the key goals of exoplanet science is the atmospheric characterisation of super-Earths.  Atmospheric abundances provide insight on the formation and evolution of those planets and help to put our own rocky planets in context.  
Observations on 55 Cancri e point towards a N-dominated atmosphere. In this paper we explore this possibility, showing which will be the most abundant gases and observable species in emission and transmission spectroscopy of such an atmosphere. 
We use analytical arguments and observed parameters to estimate the possible thermal profile of the atmosphere and test three different extreme possibilities. The chemistry is calculated using equilibrium calculations and adopting Titan's elemental abundances as a potential N-dominated atmospheric composition. We also test the effect of different N/O ratios in the atmosphere. Emission and transmission spectra are computed and showed with a resolution relevant to future missions suitable to observe super-Earths (e.g. JWST,  ARIEL). 
We find that even though N$_2$ is the most abundant molecule in the atmosphere followed by H$_2$ and CO, the transmission spectra shows strong features of NH$_3$ and HCN, and CO and HCN dominate emission spectra. We also show that a decrease in the N/O ratio leads to stronger H$_2$O, CO and CO$_2$ and weaker NH$_3$ and HCN features. A larger N/O is also more consistent with observations. 
Our exploration of a N-atmosphere for 55 Cancri e serve as a guide to understand such atmospheres and provide a reference for future observations. 
\end{abstract}

\begin{keywords}
planets and satellites: atmospheres, composition, detection
\end{keywords}

%
%________________________________________________________________

\section{Introduction}
Characterising exoplanet's atmospheres is the new frontier in exoplanet studies. The knowledge of abundances of molecular species on exoplanet atmospheres may reveal how the planet was formed and evolved \citep{ob11, alexI, nestor17,madu17}.  While many hot-Jupiters have been observed, the great observational challenges of detecting atmospheres in small, mini-gas and rocky planets makes their atmospheric properties highly unknown.  

55 Cancri e (also called Janssen) is a super-Earth of 8.703M$_{\oplus}$ and 2.023R$_{\oplus}$ \citep{cr18} located at 0.0154 AU from its host star \citep{ne14}.  Radius and mass measurements combined with interior model calculations suggest that the planet might have a high-mean-molecular-weight atmosphere \citep{de11,wi11,bo18}.  This scenario is also supported by Ly$\alpha$ observations, which show that the planet doesn't have a light H-dominated atmosphere \citep{eh12}.  The high temperatures observed (1300 - 2800K, \citet{de11}), led to the thought that it might have outgassing that enriches a primary solar-composition atmosphere \citep{ma17} or might be essentially a rock with a small heavy Na-dominated outgassed atmosphere \citep{mi11, sf09}. Nevertheless, the search for Na in its atmosphere with high-resolution spectroscopy was non-conclusive \citep{ha16}.  Other searches for molecular features using transit spectroscopy \citep{ts16} and high-resolution spectroscopy \citep{es17} turn out also with non-conclusive results.  Better constrains were obtained with recent phase curve observations \citep{de16},  and a subsequent re-analysis of these observations by \citet{hu17} and 3D calculations by \citep{pi17},  which show that the peak of the planet's radiation observed prior to the occultation, and the large day-night temperature contrast might be explained with a N-dominated atmosphere. 

Based on these last results, the aim of this paper is to explore the most relevant features in the spectra and potential observability of molecular species in a N-dominated atmosphere for 55 Cancri e.  While detailed interior composition and posterior evolution and formation of the atmosphere is largely unknown and out of the scope of this work, we study the spectral features of a N-dominated atmosphere using Titan's elemental abundances composition and explore the effect of different N/O ratios.  We calculate the most abundant elements for such atmosphere exploring different thermal profiles and model emission and transmission spectra between 3 and 20 $\mu$m to analyse the potential observability with future instrumentation such as the MIRI intrument on JWST and ARIEL.  

\section{Atmospheric Model} \label{Atmospheric Model}
\subsection{Thermal profile}\label{tp}

\begin{figure}
 \begin{center}
\includegraphics[angle=0,width=.45\textwidth]{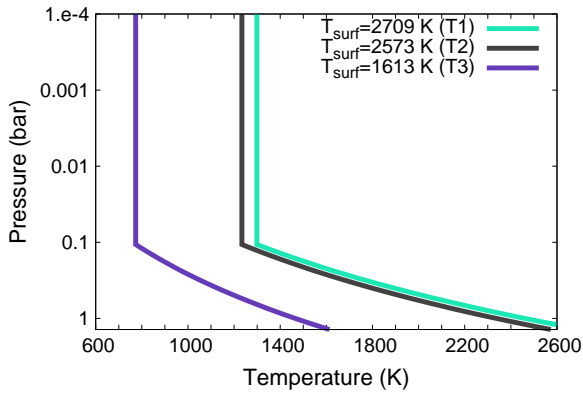}
 \end{center}
 \caption{Temperature and pressure profiles for 55 Cancri e derived using an analytic formulae and parameters from \citet{hu17} (see text). The three profiles represent extreme possibilities derived from observations, using a maximum temperature (T1, green), a mean value (T2, grey) and a minimum temperature (T3, purple).}
  \label{fig:TP}
\end{figure}

The thermal profile of 55 Cancri e is estimated using an analytical approach, where the temperature follows a dry adiabatic profile \citep{pi10} from a pressure of 1.4 bar \citep{hu17} until 0.1 bar and an isothermal profile for lower pressures, following the approach by \citet{mo17} for small rocky exoplanets.  

\begin{equation}
T(P) = 
\begin{cases}
Isothermal & \text{if } P \leq 0.1 bar,\\
T_{surf}\big(\frac{P}{1.4~bar}\big)^{\frac{R}{c_p}} & \text{if } 0.1 < P \leq 1.4 bar
\end{cases}
\end{equation}
For this calculation, we assume that the atmosphere is made of N$_2$ and for the temperature at the surface, we use the values derived in \citet{hu17}, which calculated a maximum hemisphere-averaged temperature of $\simeq$2709K (hereafter T1), an average dayside temperature of $\simeq$2573K (T2) and a minimum hemisphere-averaged temperature of $\simeq$1613K (T3).  Given that we are using an approximated thermal profile, we explore different, extreme possibilities, to explore potential changes in the spectral features due to this approximation. The three atmospheric profiles used in the paper are shown in Figure \ref{fig:TP}. 

\subsection{Atmospheric Chemistry}\label{chemistry}
Recent observations of 55 Cancri e suggest the possibility of a high-mean-molecular-weight atmosphere \citep{de16, eh12}. Furthermore, the analysis performed by \citet{hu17}, suggest that the planet might have a N$_2$-dominated atmosphere with a small percentage of H$_2$O and CO$_2$. To study this scenario, and as a proxy for a N$_2$-dominated atmosphere, we use the elemental abundances in Titan's atmosphere, where the mole fractions of each element are N=0.96287, H=0.03, C=0.007 and O=2.5e-5 \citep{mo17}.  Since the atmospheric composition of 55 Cancri e is highly unknown, in section \ref{comp} we also consider two other atmospheric compositions, keeping a N-dominated atmosphere but changing the N/O ratio.  We calculate the abundance of the gases in the atmosphere using the equilibrium chemistry code TEA, which computes the final mixing fractions using the Gibbs free-energy minimisation method \citep{bl16}. 
\begin{figure}
 \begin{center}
\includegraphics[angle=0,width=.5\textwidth]{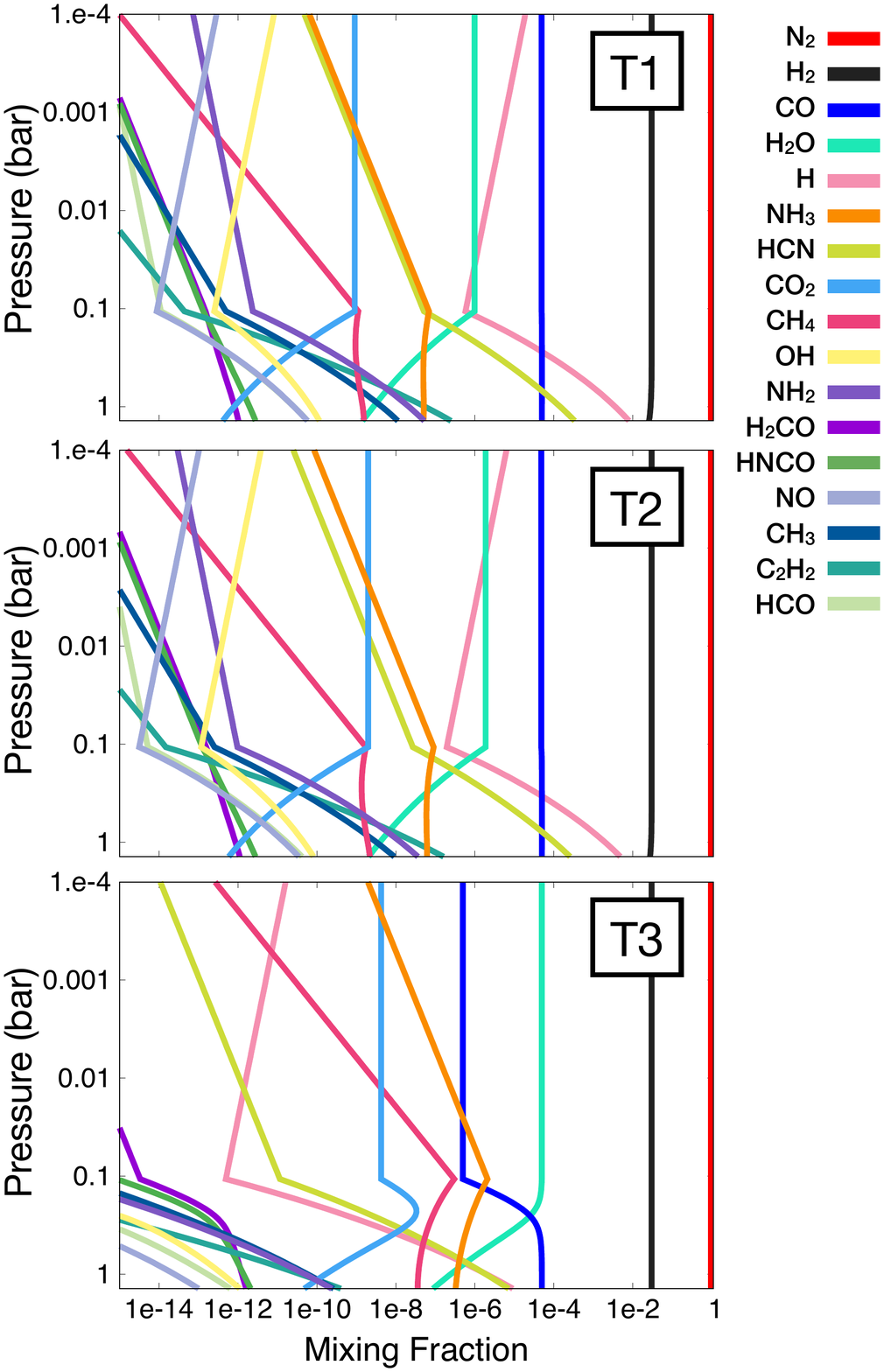}
 \end{center}
 \caption{Mixing fraction of each gas as a function of atmospheric pressure. We show results using 3 different thermal profiles as input: profile T1 (top), T2 (medium) and T3 (bottom panel).The most abundant gases are shown with different colours as indicated, where the species in the label are written in order from high to low abundance for profile T2 at 0.01 bar. }
  \label{fig:chemistry}
\end{figure}

Figure \ref{fig:chemistry} shows mixing fractions of the most abundant gases in the atmosphere. These abundances depend on the thermal profile, therefore we see differences between the three cases. For all profiles N$_2$ is the most abundant gas in the atmosphere with a mixing fraction of $\sim 0.956$, followed by H$_2$ (for all pressures). Cases T1 and T2 have a high CO abundance  close to $5 \times 10^{-5}$ (for all pressures), while for T3 CO abundance is two orders of magnitude lower ($\sim 5 \times 10^{-7}$, for P$<$0.1 bar).  T3 profile has a lower temperature that favours the formation of H$_2$O for pressures lower than 0.1 bar (mixing fraction of $\sim 5  \times 10^{-5}$), while H$_2$O abundance is lower for T1 and T2 (mixing fraction of $\sim 2  \times 10^{-6}$). Other abundant gases for N-chemistry are NH$_3$ and HCN, which are also seen in the spectra (see section \ref{results}).   
 
 \begin{figure*}
  \begin{center}
  \subfigure[]{\includegraphics[angle=0,width=.45\textwidth]{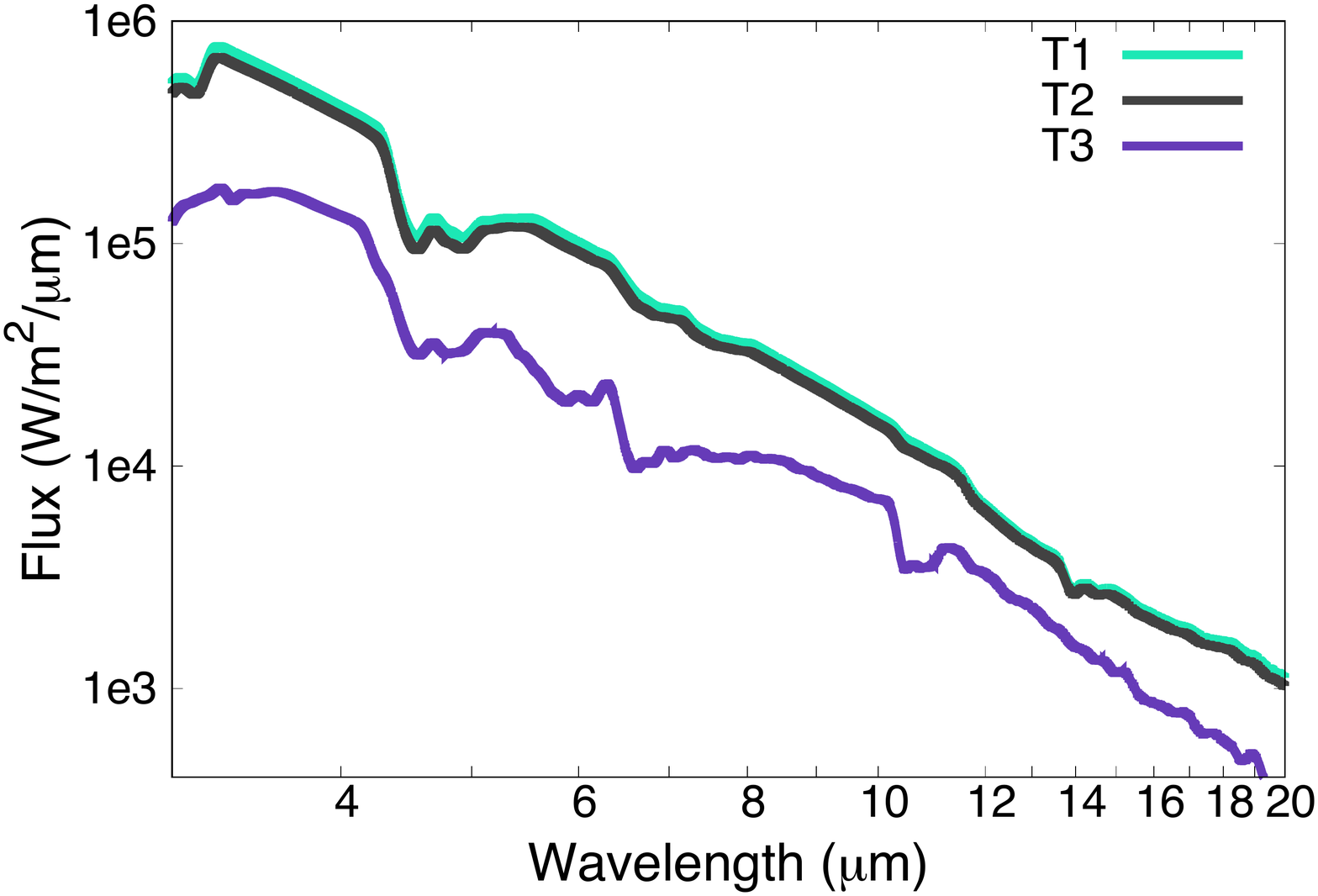}}\subfigure[]{\includegraphics[angle=0,width=.45\textwidth]{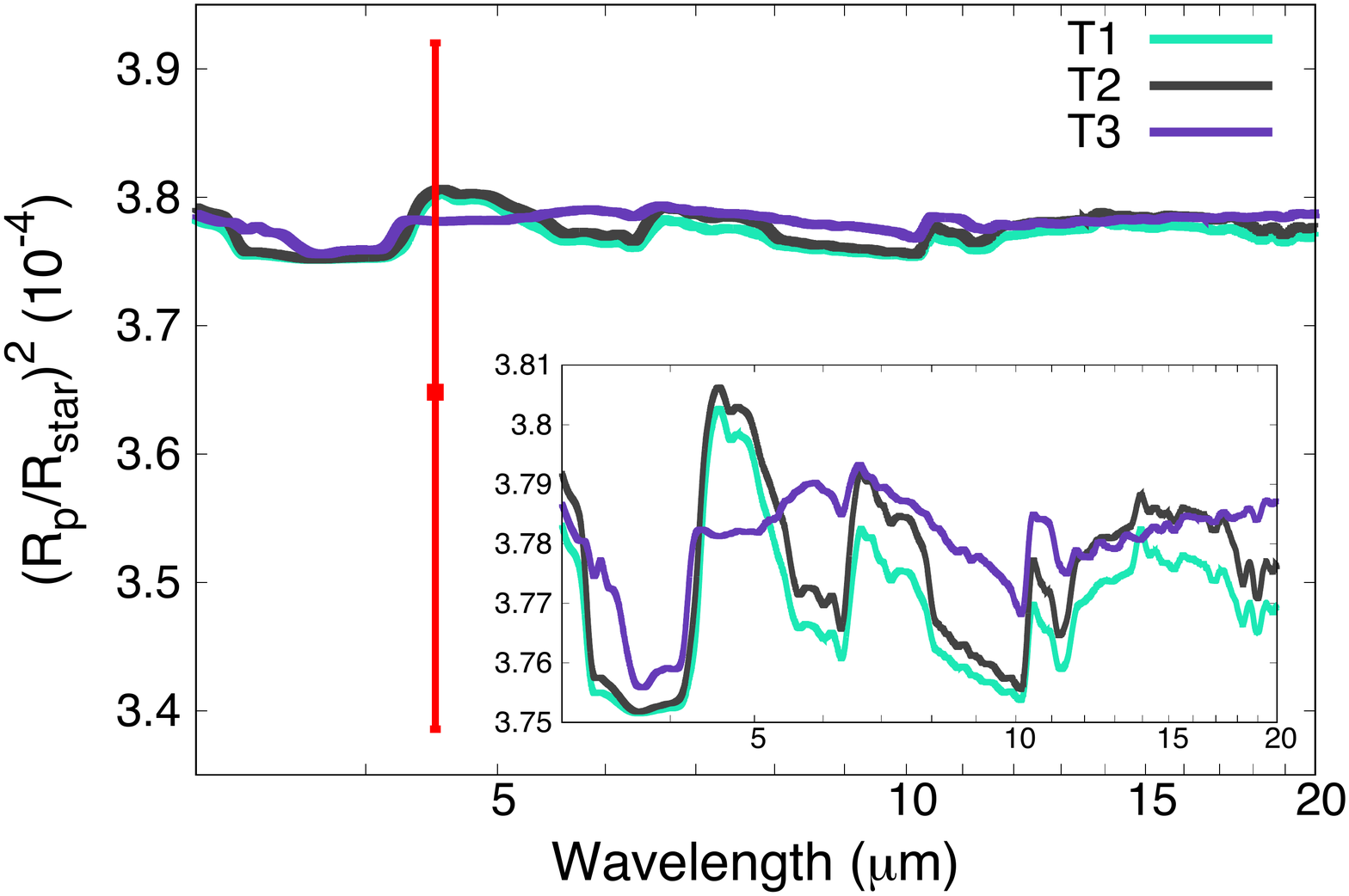}}
\end{center}
  \caption{Emission (left) and transmission (right panel) synthetic spectra of 55 Cancri e between 3 and 20 $\mu m$ for the three thermal profiles and same color scheme shown in Figure \ref{fig:TP}.  The red point in the right panel shows the average transit depth of 55 Cnc e observed with Spitzer \citep{de16}.}
  \label{fig:spectra}
\end{figure*}

\subsection{Synthetic Spectra}
The synthetic spectra of 55 Cancri e are calculated using the Smithsonian Astrophysical Observatory 1998 (SAO98) line-by-line radiative transfer code, originally built for Earth like planets \citep{ts76, tj02, k07} and subsequently tested and used in multiple papers for rocky exoplanets \citep{kt09, r13,ru18} and recently adapted for hot-Jupiter modelling \citep{w17}.  We calculate emission and transmission spectra at high resolution (0.1$~cm^{-1}$ wavenumber steps) and smear them out to a resolving power of 100 to simulate the resolution that we will obtain with the MIRI instrument on JWST, \citep{gr16} and relevant for other future low resolution instrumentation (e.g.  ARIEL, \citet{ARIEL}). Important molecules such as H$_2$O, CH$_4$, CO, CO$_2$, NH$_3$, N$_2$, HCN, H$_2$, O, O$_3$, O$_2$, H$_2$CO, C$_2$H$_2$, N$_2$O$_5$, N$_2$O, NO$_2$ are considered with linelists from HITRAN (Rothman et al. 2013) and HITEMP (Rothman et al. 2010) databases.

\section{Results}\label{results}
We model emission and transmission spectra for the three thermal profiles described in section \ref{tp} and using Titan's elemental abundances (section \ref{chemistry}). The results are shown in Figure \ref{fig:spectra}. While the difference in the isothermal part of profiles T1 and T2 is only $\sim$65 K, the difference between T2 and T3 is $\sim$460 K, which causes large differences in the abundances (Figure \ref{fig:chemistry}) and are also seen in the spectra (Figure \ref{fig:spectra}). For this reason, profiles T1 and T2 have similar features in emission and transmission, while T3 looks different.  

\begin{figure*}
  \begin{center}
\includegraphics[angle=0,width=0.8\textwidth]{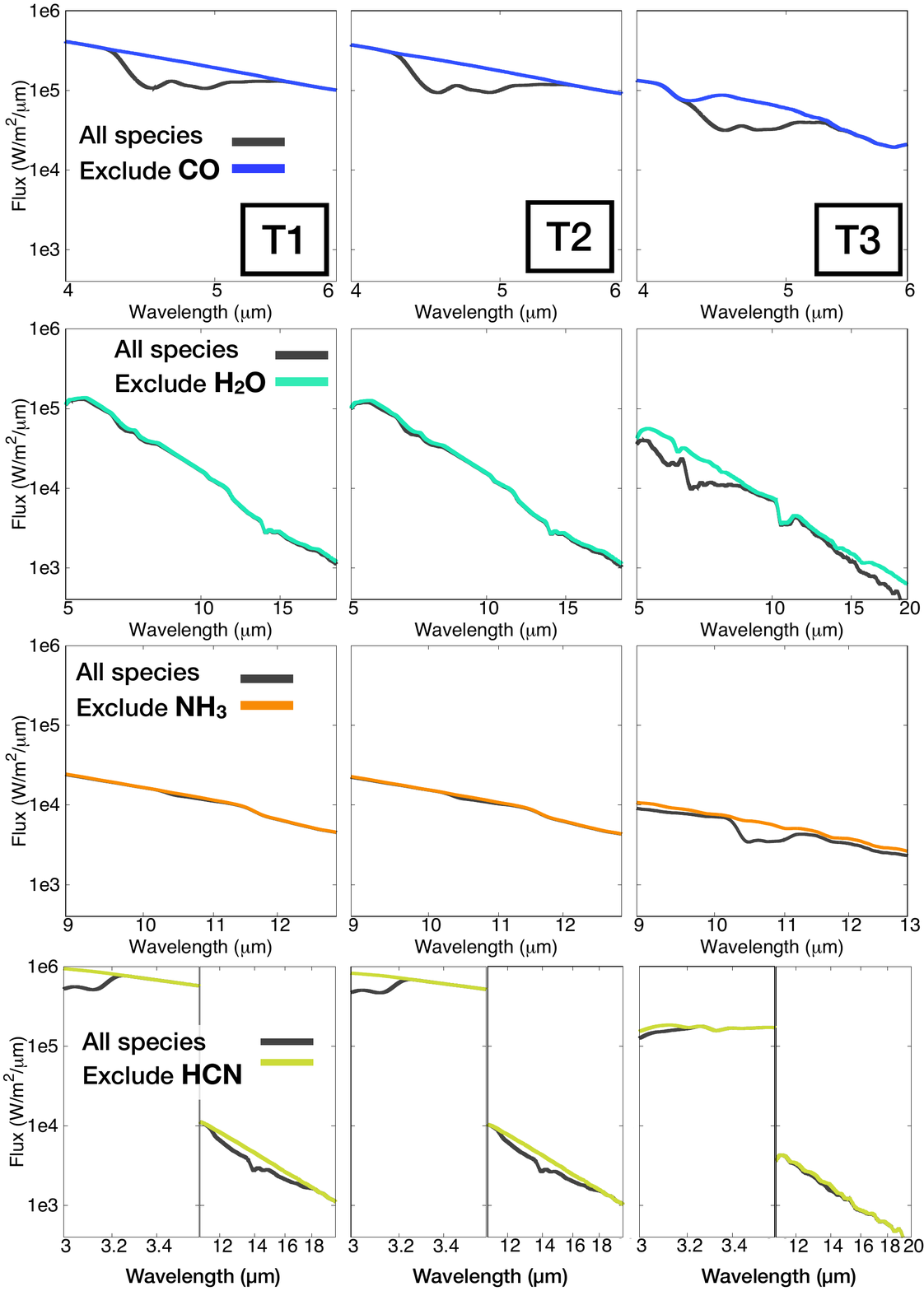}
 \end{center}
  \caption{Spectral features in emission for the most relevant species and for the 3 different cases: profile T1 (left), T2 (middle) and T3 (right column).  Spectra including all species is shown in grey and spectra with all species but excluding one of them is shown in colours (same colour code as in Figure \ref{fig:chemistry}). Note that in order to highlight the individual features of different species, we show a different wavelength range in the different rows. HCN is divided in two to show its two relevant features. }
  \label{fig:emission}
\end{figure*}

\begin{figure*}
  \begin{center}
\subfigure[]{\label{max}\includegraphics[angle=0,width=.33\textwidth]{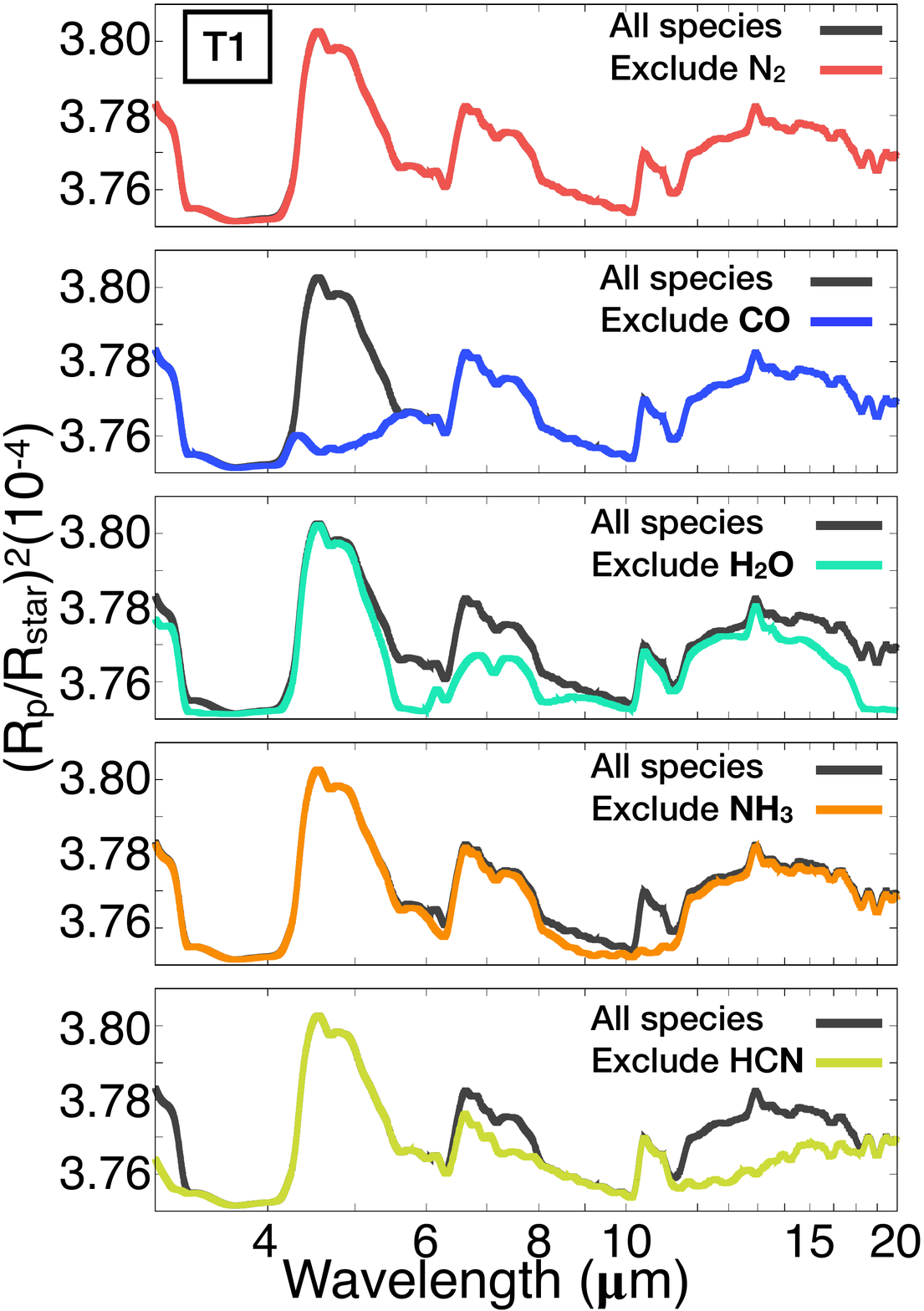}}\subfigure[]{\label{modeled}\includegraphics[angle=0,width=.33\textwidth]{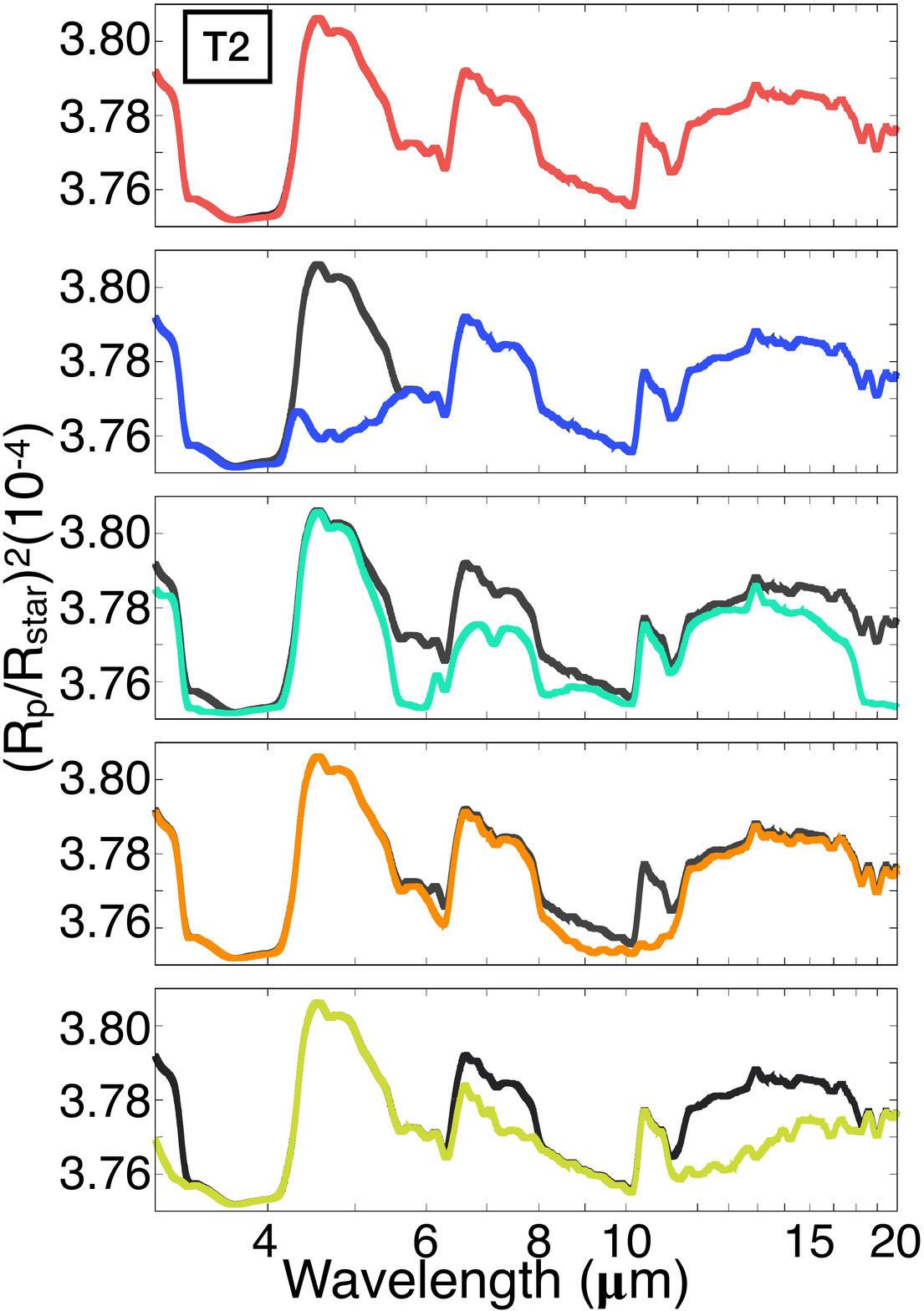}}\subfigure[]{\label{min}\includegraphics[angle=0,width=.33\textwidth]{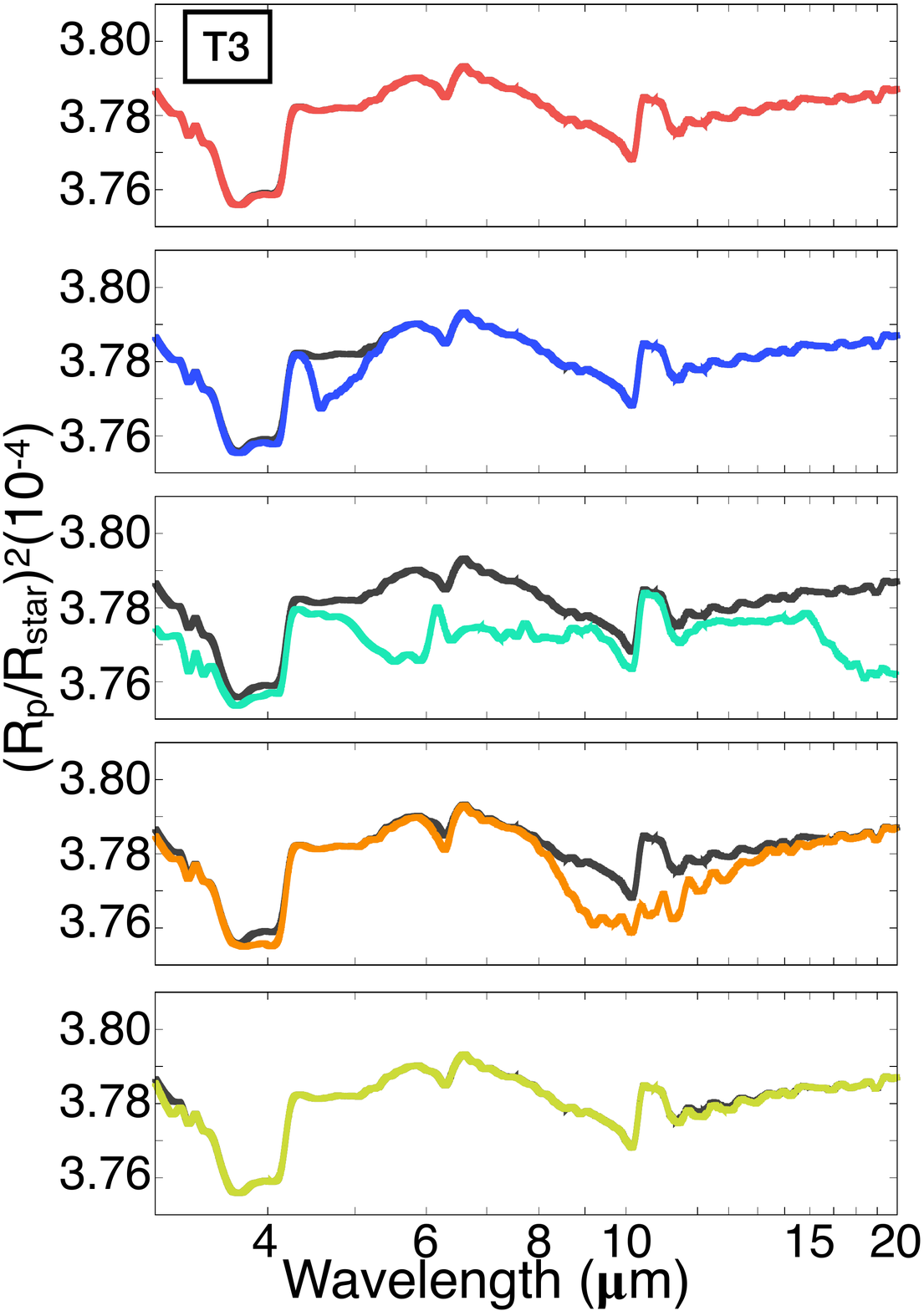}}
 \end{center}
  \caption{Transmission synthetic spectra for profile T1 (left), T2 (middle) and T3 (right panel).  Same color code as in Figure \ref{fig:emission}. }
  \label{fig:transit}
\end{figure*}

In order to understand the influence of individual species in the spectra,  we calculate synthetic spectra with all the species and with all the species except one of them. This is shown in Figure \ref{fig:emission} for emission and in Figure \ref{fig:transit} for transmission.

We show in Figure \ref{fig:emission} that emission spectra present a prominent feature of CO at 4.5-5.5 $\mu$m for all three profiles. H$_2$O features are not relevant in the two hottest cases (T1 and T2) but it has strong features at 5-9 and 13-20 $\mu$m in the coldest case (T3). Regarding N-bearing species, NH$_3$ has features at 10-11 $\mu$m only in the coldest case (T3), while HCN features at 3-3.5 and 11-18 $\mu$m are more pronounced in the two hottest cases (T1 and T2).  

As seen in Figure \ref{fig:transit},  we also see strong CO features at 4.5-5.5 $\mu$m in transmission spectra.  We see NH$_3$ features at 5.5-6.1 and 8-11 $\mu$m, although the former feature is hidden by the stronger H$_2$O feature at the same wavelength. 
HCN features are strong at 3.2, 6.1-8 and 11-18 $\mu$m. H$_2$O has strong features at 5-9 and 14-20 $\mu$m, although the features at 7-8 and 14-18 $\mu$m are hidden by HCN.  For the T3 profile, H$_2$O features at 3-3.8, 4.5-8,5 and 13-20 $\mu$m dominate the transmission spectra, with the exception of NH$_3$ and CO features at 8-13 and 4.5-5.2 $\mu$m, correspondingly.  

\begin{figure}
 \begin{center}
\includegraphics[angle=0,width=.45\textwidth]{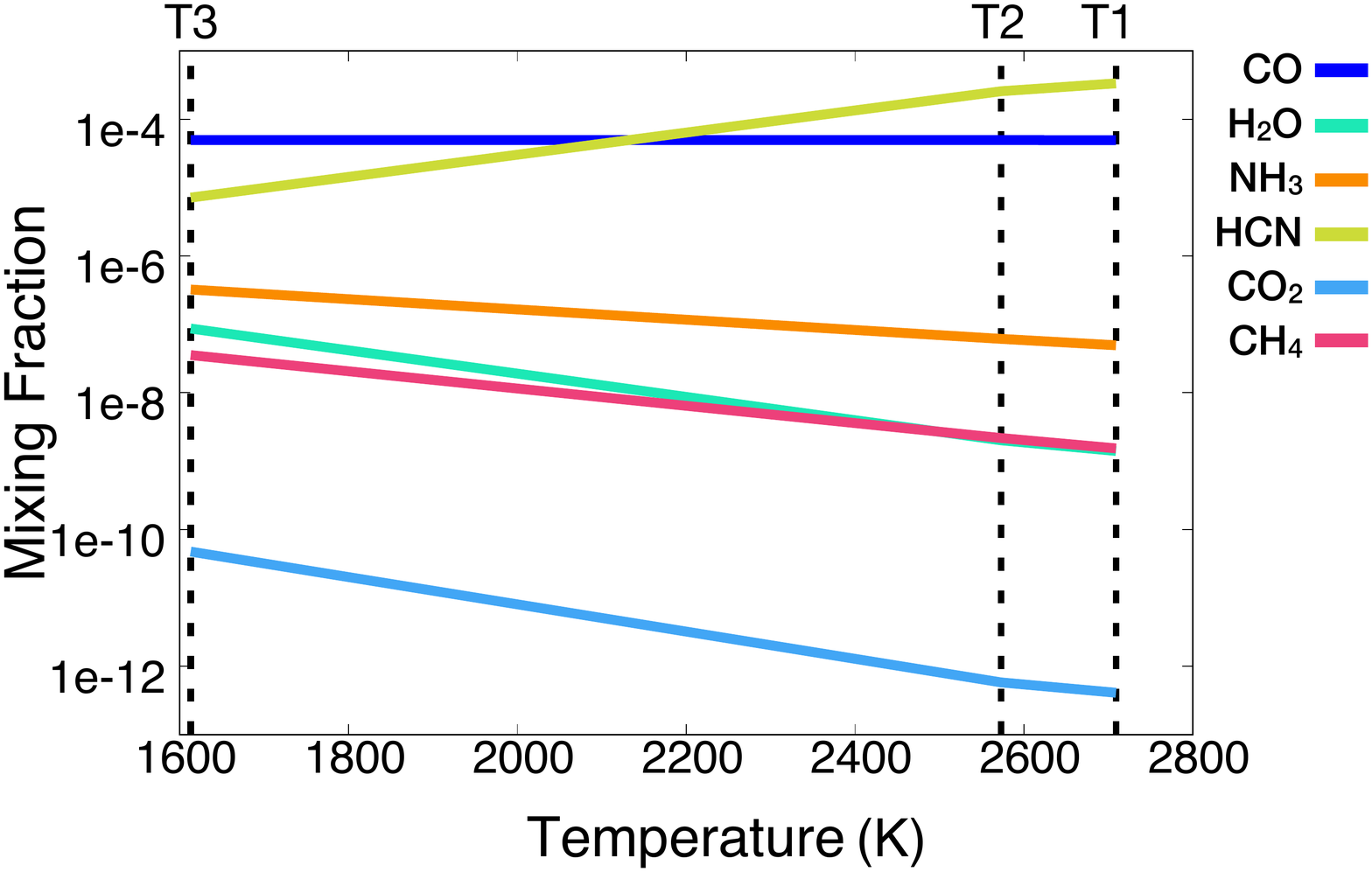}
 \end{center}
 \caption{Mixing fraction of the most abundant observable molecules in the atmosphere (CO, H$_2$O, NH$_3$, HCN, CO$_2$,CH$_4$) as a function of temperature at a pressure of 1.4 bar.}
  \label{fig:TX}
\end{figure}

We are probing deep in the atmosphere of 55 Cnc e, at a pressure of $\sim 1.4$ bar \citep{hu17}. Figure \ref{fig:TX} shows the concentration of the most abundant observable molecules in the atmosphere as a function of the temperature, for a pressure of P=1.4 bar.  The figure shows that HCN is the most abundant molecule for T$>$2100 K, which explains the strong HCN features observed both in emission and transmission for T1 and T2. CO is the second most abundant species for T1 and T2, and also presents strong features. On the other hand, H$_2$O is not so abundant for T1 and T2, but it becomes more abundant for T3, where we observe the strongest H$_2$O features. Nevertheless, in all cases in transmission (T1, T2 and T3) and in the coldest case (T3) in emission we observe H$_2$O features. This is because H$_2$O is a very strong intrinsic absorber and one of the most important sources for the opacity in the infrared, even if its not the most abundant molecule in the atmosphere. 

N$_2$ is the most abundant gas in the atmosphere, but it shows no spectral features in transmission or emission, being HCN and NH$_3$ the N-carriers that show the presence of a N-dominated atmosphere for the three different temperature profiles explored.  

\subsection{The effect of different atmospheric composition}\label{comp}
The origin and atmospheric composition of rocky super-Earths is highly unknown.  In the solar system, the atmospheric composition of rocky bodies is a consequence of accretion of material at early ages and the subsequent outgassing and escape of volatiles during the planet's evolution. One of the parameters that can provide information to trace the history of these bodies is the N/O ratio. In protoplanetary disks, N/O ratio depends on the depletion of N$_2$ and O-carriers (H$_2$O, CO$_2$, CO) and it increases when each of these O-bearing species freeze and the ice-lines are crossed \citep{pi16}.  A good example of the relevance of these processes to understand the formation history of the planet is Titan, where it is believed that the present levels of N$_2$ are a result of accretion of primordial NH$_3$ ices present in the protosolar nebula \citep{at78, ma14}.  For this reason, and due to the high uncertainties in the composition of 55 Cancri e's atmosphere, we explore different N/O ratios maintaining a N-dominated atmosphere, to explore the changes in the chemistry and asses possible observability with future instrumentation.  

\begin{figure}
 \begin{center}
\includegraphics[angle=0,width=.45\textwidth]{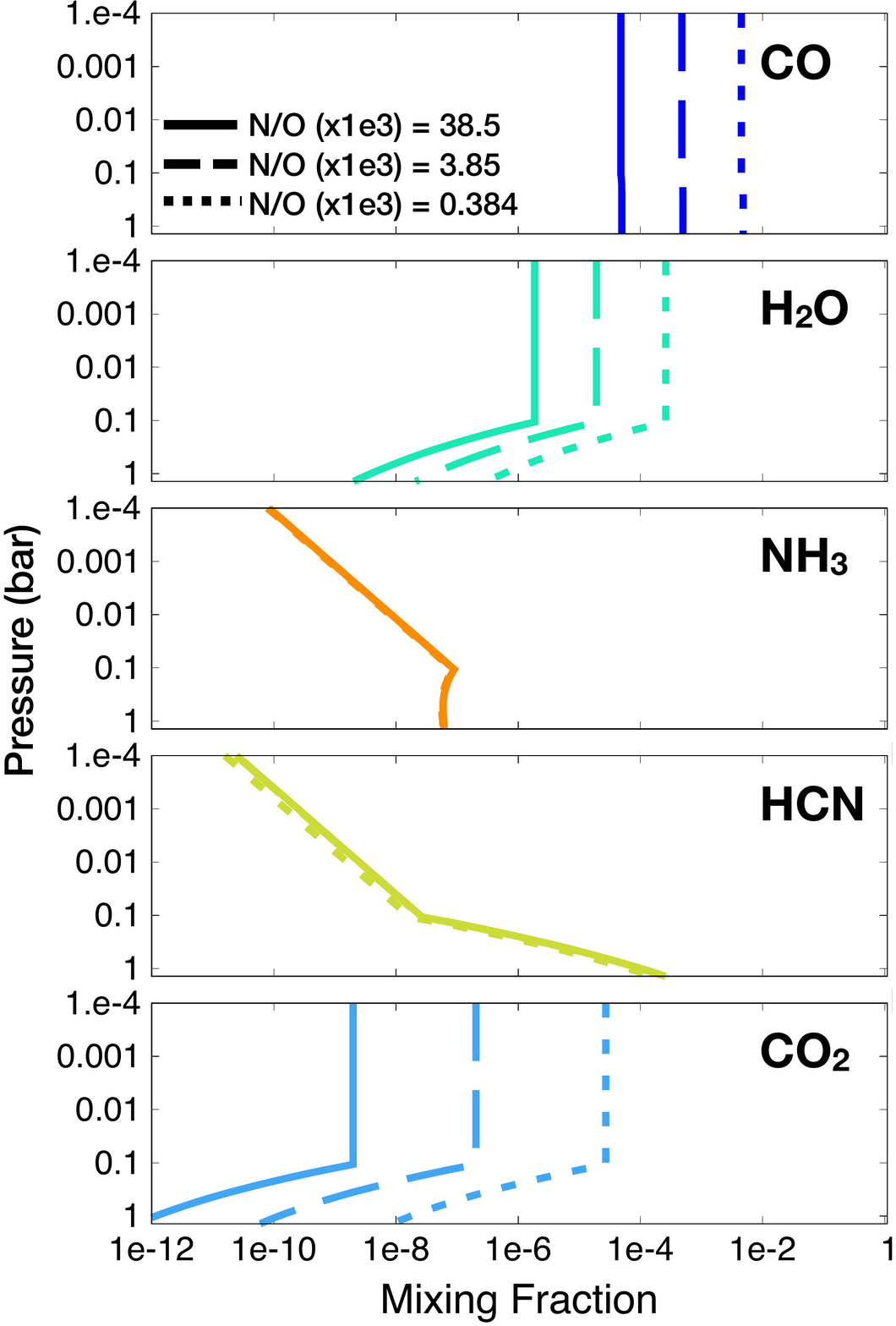}
 \end{center}
 \caption{Mixing fractions of some relevant O- and N-carriers in the atmosphere. Different panels show different species and different lines represent the 3 different N/O ratios explored. }
  \label{composition}
\end{figure}

In addition to Titan's N/O = 3.85$\times10^4$ (see section \ref{chemistry}), we also explore two other cases: N/O = 3.85$\times10^3$ and N/O = 3.85$\times10^2$, maintaining Titan's abundances for C and H.  In these two additional cases we use profile T2. Figure \ref{composition} shows the change in the atmospheric chemistry.  We see that an increase in O leads to a larger concentration of CO, H$_2$O, and specially CO$_2$, which increases $\sim$2 orders of magnitude. On the other hand, even though the percentage of N is lower, we don't see any significative change in NH$_3$ and HCN concentrations. 

\begin{figure*}
  \begin{center}
  \subfigure[]{\includegraphics[angle=0,width=.45\textwidth]{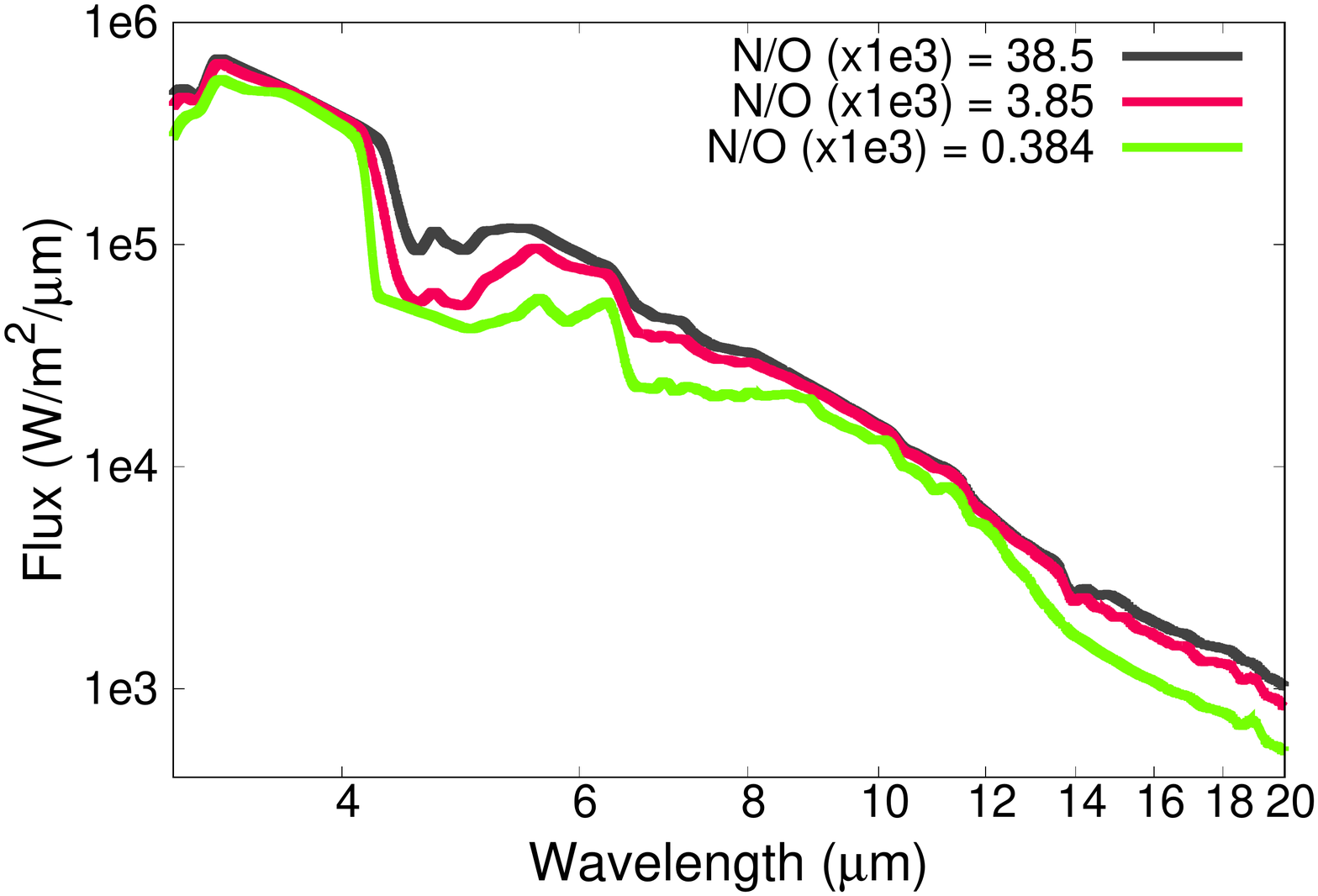}}\subfigure[]{\includegraphics[angle=0,width=.45\textwidth]{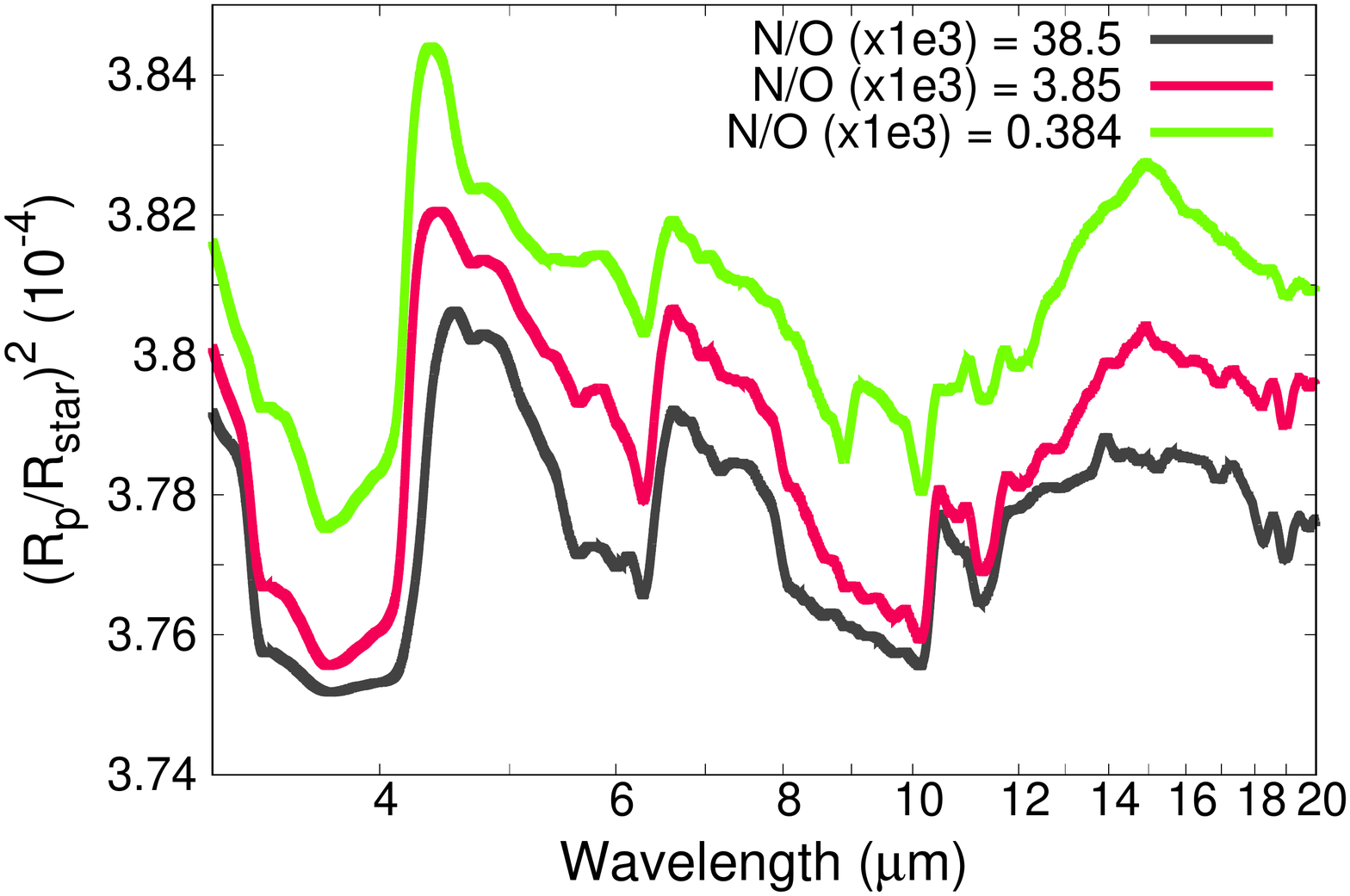}}
\end{center}
  \caption{Emission (left) and transmission (right panel) spectra calculated using three different N/O ratios in the atmosphere.}
  \label{fig:NOspectra}
\end{figure*}

\begin{figure*}
  \begin{center}
\includegraphics[angle=0,width=0.9\textwidth]{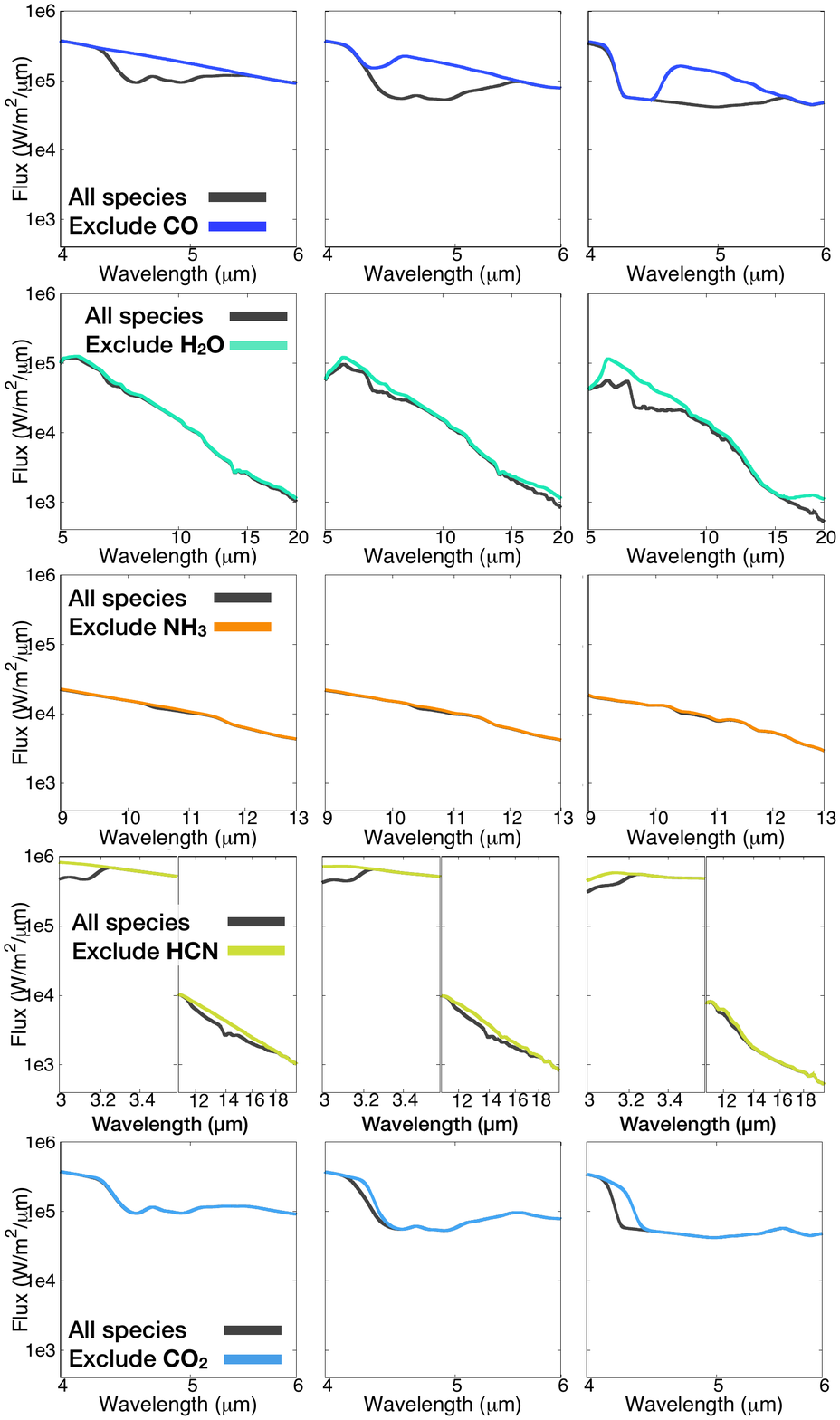}
 \end{center}
  \caption{Effect of different species in the emission spectra and how they change with different N/O ratios. We show results with N/O = 3.85$\times10^4$ (left),  N/O = 3.85$\times10^3$ (center) and N/O = 3.85$\times10^2$ (right column). Same color code as in Figure \ref{fig:emission}.}
  \label{fig:emission-comp}
\end{figure*}

\begin{figure*}
  \begin{center}
\subfigure[]{\label{min}\includegraphics[angle=0,width=.33\textwidth]{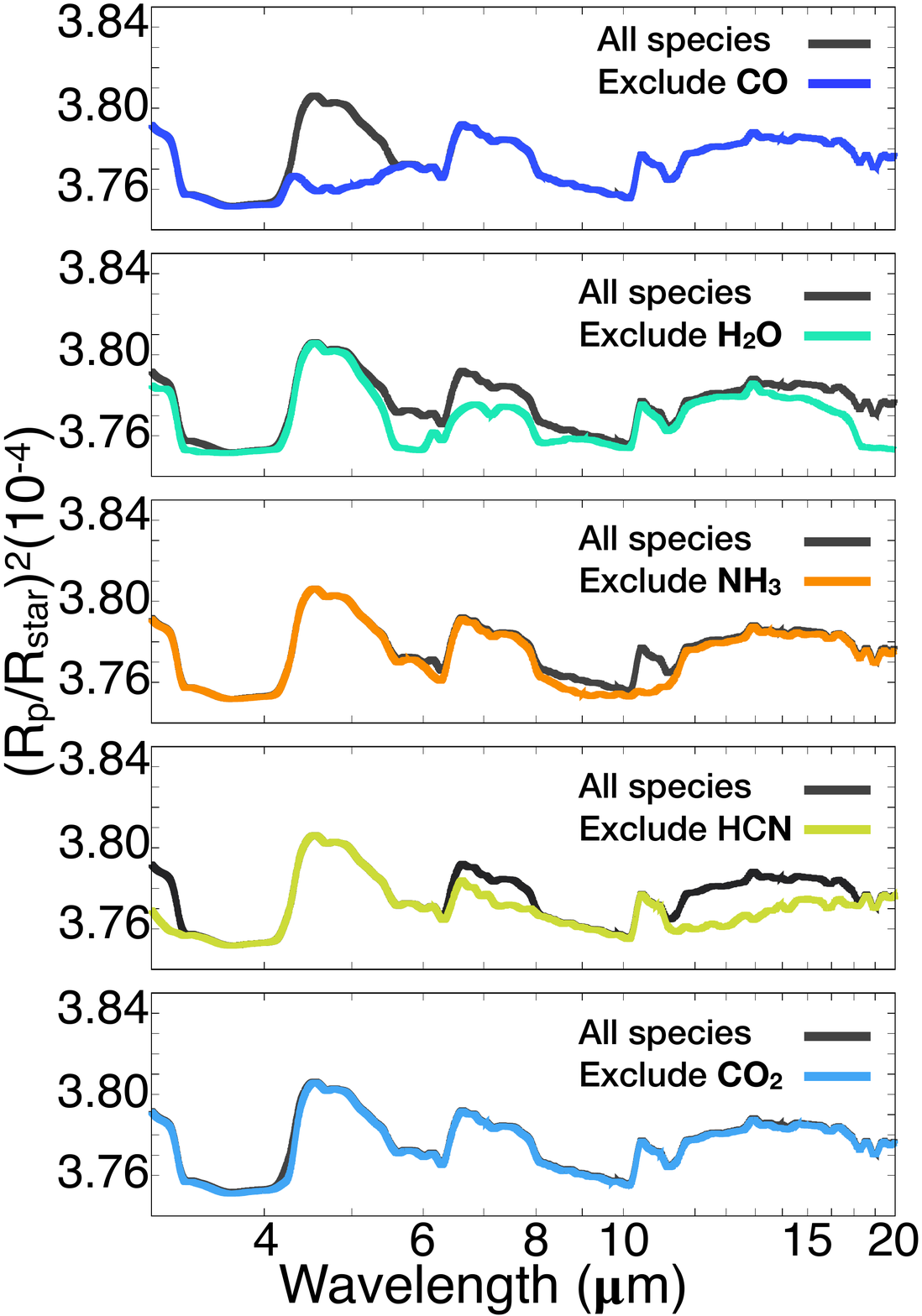}} \subfigure[]{\label{modeled}\includegraphics[angle=0,width=.33\textwidth]{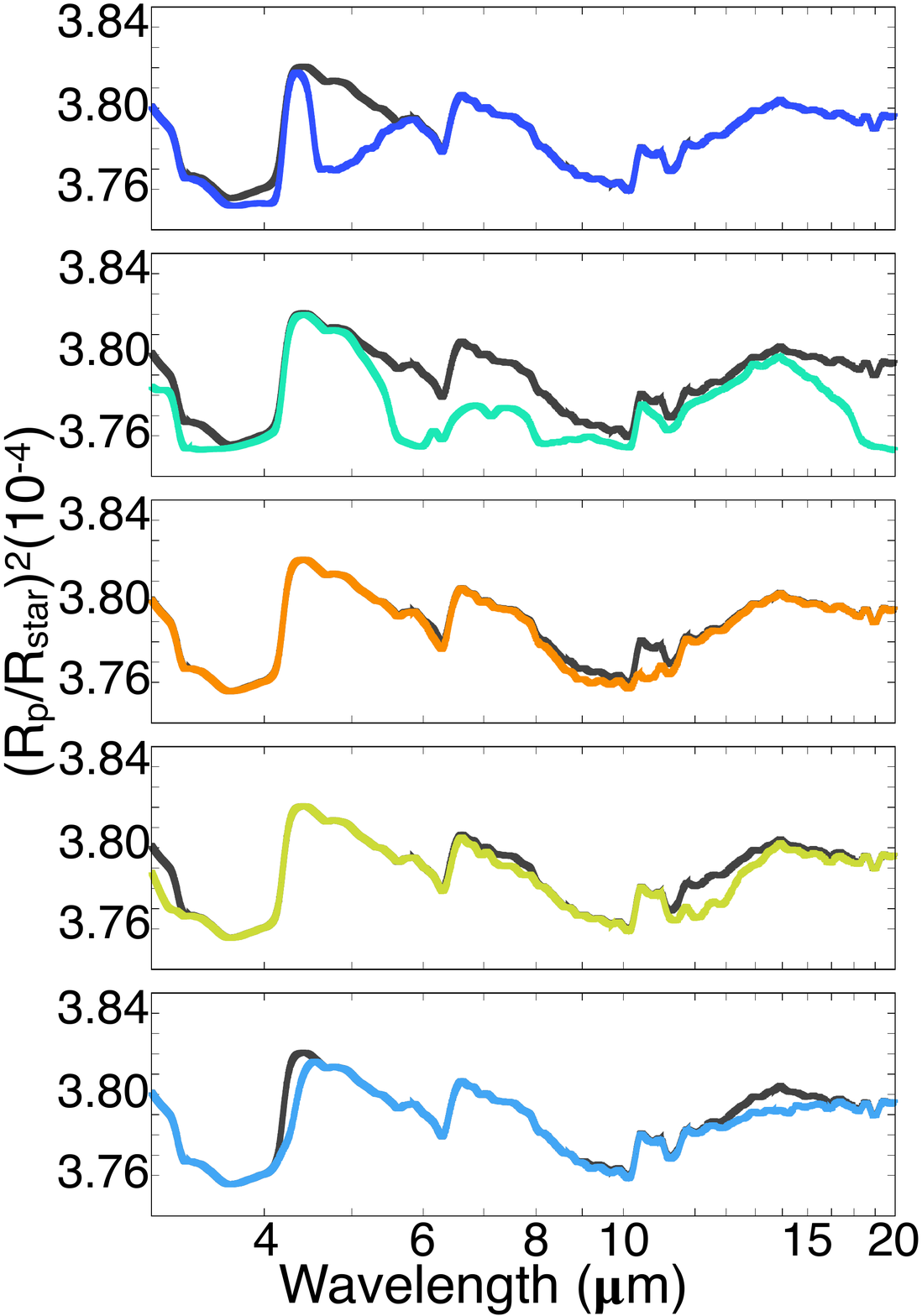}}\subfigure[]{\label{max}\includegraphics[angle=0,width=.33\textwidth]{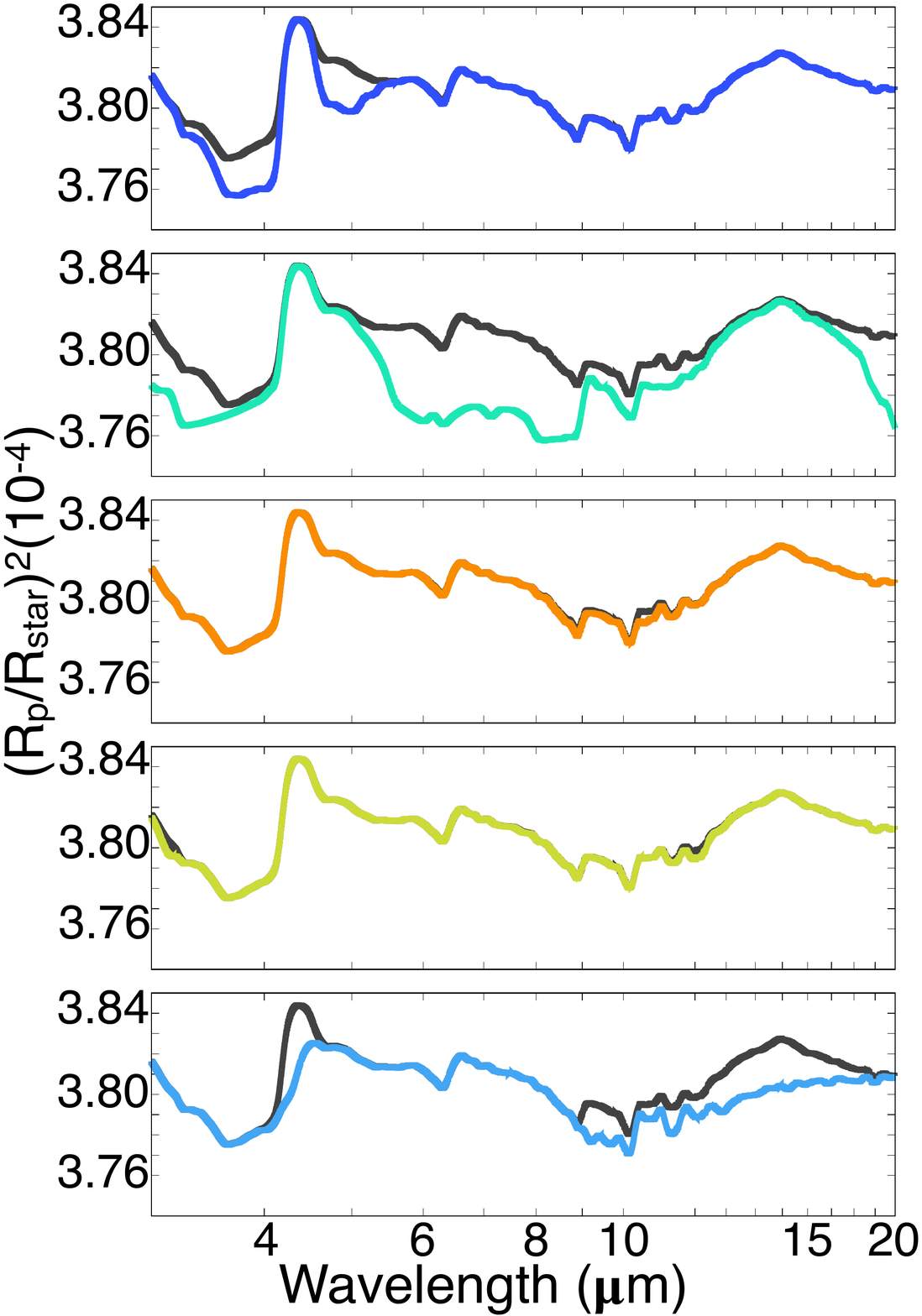}}
 \end{center}
  \caption{Consequence of a change in the N/O ratio in the transmission spectra.  We show spectra adopting N/O = 3.85$\times10^4$ (left),  N/O = 3.85$\times10^3$ (center) and N/O = 3.85$\times10^2$ (right panel) in the atmosphere.  The spectra including all the species are shown in grey and the spectra excluding a specific molecule (indicated in the figure on the left panel, same for the other panels) is shown in colours. }
  \label{fig:transit-comp}
\end{figure*}

These differences in concentrations are seen in the spectra (Figure \ref{fig:NOspectra}). The relevance of each individual species in the spectra is shown in Figure \ref{fig:emission-comp} for emission and in Figure \ref{fig:transit-comp} for transmission spectra. 

Figure \ref{fig:emission-comp} shows that CO features grow in strength when decreasing N/O in the emission spectra. The same happens with H$_2$O and CO$_2$, which was not even relevant for the largest N/O case that we considered in section \ref{results}.  NH$_3$ has no features in the three cases, while HCN features weaken when decreasing N/O. 

For transmission, we see in Figure \ref{fig:transit-comp} that a decrease in N/O leads to stronger H$_2$O and CO features. CO$_2$ features were not observable in the transmission spectra of Titan's N/O, but they become noticeable as we decrease the N/O.  On the other hand, we see weaker NH$_3$ and HCN features when decreasing the N/O. Interestingly,  NH$_3$ and HCN vanish in the case with the smallest N/O, implying that if O levels are high enough, we can not observe any N-carrier in the transmission spectra in a N-dominated atmosphere.  We also see that the R$_p$/R$_{star}$ increases as N/O decreases, therefore a large N/O is more consistent with phase curve observations \citep{hu17}.

\section {Discussion}
\subsection{Disequilibrium chemistry}
The abundances of different species in 55 Cancri e atmosphere were calculated assuming that the atmosphere is in chemical equilibrium, which is usually a good approximation for extremely hot planets \citep{li13, mi14,ve18}.  Detailed calculations including disequilibrium processes are beyond the scope of this paper and will be included in a future study. We introduce this analysis to discuss its main effects and how this might affect our calculations 

\begin{figure}
 \begin{center}
\includegraphics[angle=0,width=.45\textwidth]{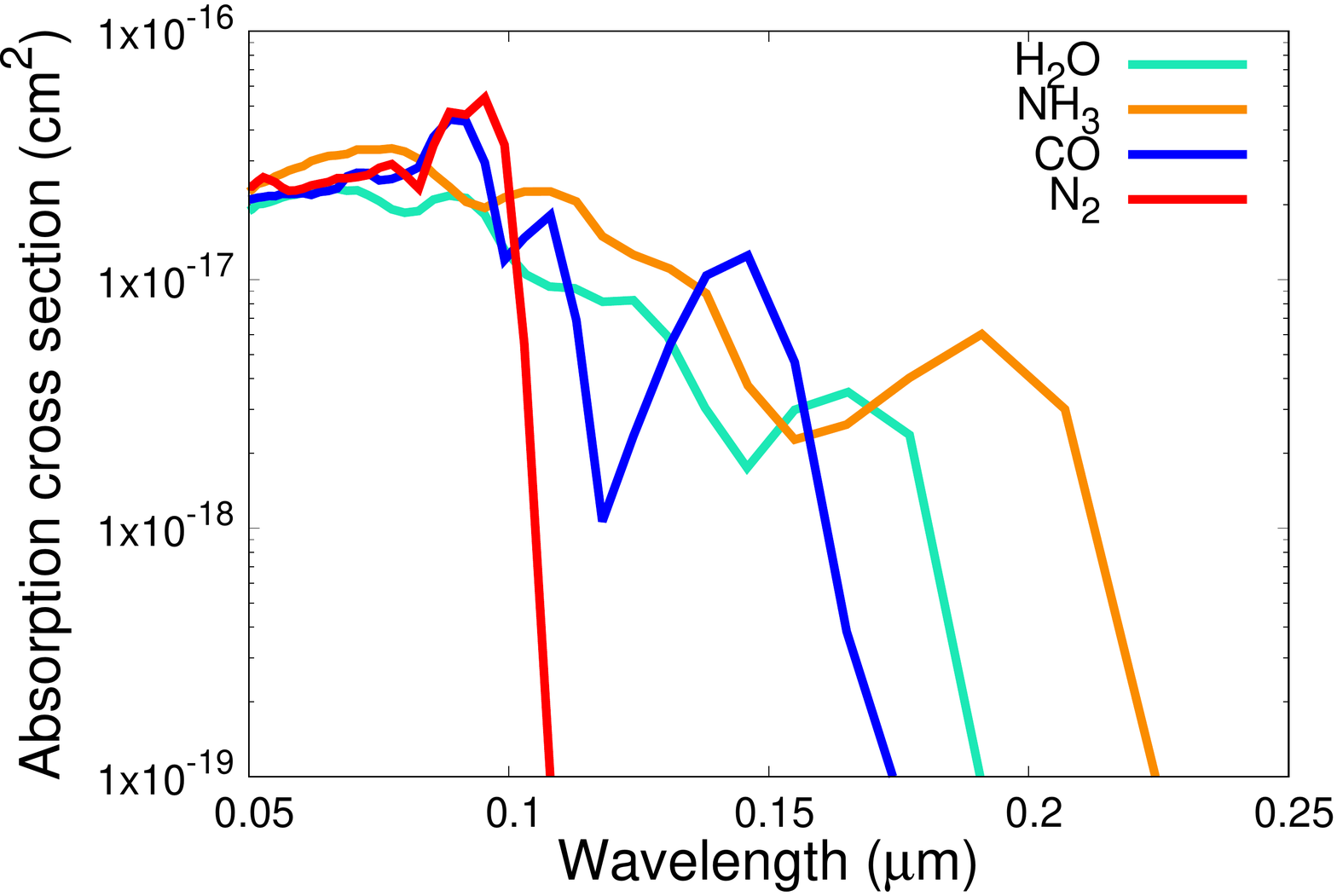}
 \end{center}
 \caption{Absorption cross sections of some of the most relevant molecules in the atmosphere: N$_2$ \citep{chN2}, NH$_3$ \citep{bu93}, CO \citep{chCO} and H$_2$O \citep{chH2O}.}
  \label{cross-section}
\end{figure}

Due to the close proximity between the planet and the host star, dissociation of atmospheric species is probably one of the most important disequilibrium proceses that occur in this atmosphere.  This process requieres an energy above the dissociation energy for each molecule, and the effectiveness of this process will also depend on the absorption cross section of each molecule at those energies. Figure \ref{cross-section} shows the absorption cross sections of some of the most relevant species in the atmosphere.  
N$_2$ is a very stable molecule, its dissociation threshold is around 0.1265$\mu$m \citep{yu99} and as seen in Figure \ref{cross-section},
 N$_2$ has a large cross section for wavelengths $\lambda < 0.1 \mu$m. This energy is probably absorbed higher in the atmosphere, so is likely that this process plays no important role in the chemistry at P$<1\times10^{-4}$.  The other N-carrier of relevance is NH$_3$, this molecule has a dissociation threshold at 0.2637 $\mu$m \citep{yu99} and its absorption is important for $\lambda \sim 0.2 \mu$m, where it has no other competitor, so this molecule will probably dissociate and form NH$_2 +$H.  In Figure \ref{fig:transit} we show how the spectra would look like without NH$_3$ being present in the atmosphere, although we note that its dissociation might affect other molecules as well. Finally, we have CO and H$_2$O. CO has a very strong bond with a dissociation energy at 0.1113$\mu$m \citep{yu99}, similar to the case of N$_2$, this energy is likely to be absorbed higher in the atmosphere. H$_2$O is not so abundant (see Figure \ref{fig:chemistry}), but since its dissociation threshold is at  0.2398$\mu$m \citep{yu99},  it is likely to dissociate in this atmosphere for $\lambda \sim 0.17 \mu$m, Figure \ref{fig:transit} shows the spectra with no H$_2$O.   
 
Including the effects of photochemistry might be important for understanding the chemistry and its effects on the spectra of hot rocky exoplanet atmospheres, specially for planets around active stars that emit strongly in the UV. 

\section{Conclusion}
Observations of 55 Cancri e suggest the presence of a high-mean-molecular-weight atmosphere \citep{de16, eh12}. A subsequent analysis of those observations and 3D modelling further support the idea of a N-dominated atmosphere \citep{hu17,pi17}. In this context, this paper makes a first approximation to explore a N-dominated atmosphere for 55 Cancri e and study its most relevant molecules and spectral features for transmission and emission spectroscopy in views of future observations (JWST,  ARIEL).  

Since hot super-Earths atmospheres are largely unknown, we use three different extreme atmospheric thermal profiles, calculated using an analytical approach and parameters derived from observations \citep{cr18,hu17}. We adopt as an example of a N-atmosphere Titan's elemental abundances and calculate the chemistry using equilibrium chemistry.  We also test the effect of changing the elemental abundances composition decreasing the N/O ratio. Finally, we compute transmission and emission spectra to a low resolution comparable to what we will be able to obtain with future instrumentation.  

Our results show that  N$_2$ is the most abundant molecule in the atmosphere for all pressures, followed by H$_2$, CO and H$_2$O, where CO is more relevant for the hottest profiles and H$_2$O for the coldest case (for P$<$0.1 bar).  
The effect of decreasing N/O ratio is an increase in H$_2$O, CO and CO$_2$ and a decrease in NH$_3$ and HCN abundances. 

Both emission and transmission spectra show strong CO features at 4.5-5.5 $\mu$m for the three thermal profiles.  H$_2$O features are also visible at 5.5-6.1 and 18-20$\mu$m for all cases considered in transmission and for the coldest case (T3) in emission. Regarding N-bearing species,  N$_2$ shows no features, but other N-bearing species, not so abundant but strong absorbers, can be seen in the spectra.  These are NH$_3$ and HCN, both with strong features. NH$_3$ features are at 8-11 $\mu$m for the three profiles in transmission and for the coldest profile (T3) in emission. HCN features are at 3, 6.1-8 and 11-18 $\mu$m, although are only strong for the two hot cases both in transmission and emission.  When decreasing the N/O, we observe stronger features of H$_2$O, CO and CO$_2$ and weaker features of NH$_3$ and HCN.  The case with the smallest N/O ratio shows no N-carrier molecules in transmission spectra and weak HCN lines in emission. We also find that a large N/O ratio is more consistent with observations. 

Our results can be used as a guide to understand what to expect in a N-dominated atmosphere for 55 Cancri e and as a reference in preparation for future observations.

\end{document}